\documentstyle[12pt]{article}
\begin{document}

\begin{titlepage}

\title{\begin{flushright}
{\normalsize UB-ECM-PF 95/13} \\
{\normalsize LGCR 95/06/05} \\
{\normalsize DAMTP R95/35}
\end{flushright}
\vspace{2cm}
{\Large \bf Limits on Kaluza-Klein Models from
COBE Results}}

\author{Yu. Kubyshin \thanks{On leave of absence from Nuclear
Physics Institute,
Moscow State University, 117234 Moscow, Russia.}
\thanks{E-mail address: kubyshin@ecm.ub.es} \\
Department d'ECM, Universitat de Barcelona, \\
Av. Diagonal 647, 08028 Barcelona, Spain  \\
 and \\
 J\'er\^ome Martin \thanks{On leave of absence from
Laboratoire de Gravitation et Cosmologies Relativistes,
Universit\'e Pierre et Marie Curie, CNRS/URA 769, Tour 22/12,
Boite courrier 142, 4 place Jussieu 75252 Paris cedex 05, France.}
\thanks{E-mail address: J.Martin@damtp.cam.ac.uk (after the 1st of July:
jmartin@ccr.jussieu.fr) } \\
Department of Applied Mathematics and Theoretical
Physics, \\
University of Cambridge, Silver street, \\
Cambridge CB3 9EW, United Kingdom}

\date{June 30, 1995}

\maketitle

\begin{abstract}

The large-angular-scale anisotropy of the cosmic microwave
background radiation
in multidimensional cosmological models (Kaluza-Klein
models) is studied. Limits on parameters of the models
imposed by the experimental data are obtained. It is shown that in
principle there is a room for Kaluza-Klein models as possible candidates
for the description of the Early Universe. However, the
obtained limits are very restrictive and none of the concrete models,
analyzed in the article, satisfy them.

\end{abstract}
\end{titlepage}


It was shown in the literature \cite{Gr-PR93} that a certain
(although quite narrow)
class of four-dimensional inflationary scenarios agrees with the
observational data \cite{cobe} on the anisotropy of the cosmic
microwave background radiation. However, many of the models
which provide mechanisms of the inflationary expansion of the
three-dimensional spatial part of the Universe arise from
string or supergravity models (see, for example, \cite{Duff})
formulated in the spacetime with more than four dimensions.
As it is known the dynamics of the multidimensional Universe
differs significantly from that of the four-dimensional one
at the Early stage of the evolution, when the extra dimensions
could have an important role to play. In particular, this is
reflected in the spectrum of gravitational waves.
The reason is basically due to the
fact that, apparently, the scale factor of the space of extra dimensions
was not static and was comparable to the scale of the three-dimensional
spatial part at that epoch. This suggests that it is important
to derive constraints on the models of the Early Universe within the
framework of multidimensional (Kaluza-Klein) cosmology.

Our discussion in this article will be rather general in order
to include a wider class of models, not just those which stem from
superstring or supergravity theories. We expect that recent discovery
of the angular variation in the temperature of the cosmic microwave
background radiation (CMBR) by the Cosmic Background Explorer (COBE)
will definitely shed light on the origin and nature of long-wavelength
cosmological perturbations and in this way may give some evidence
{\em pro} or {\em contra} the Kaluza-Klein hypothesis.
In this letter we study gravitational waves generated quantum
mechanically and calculate the temperature variation of the CMBR.
Then we combine the result of this calculation, results of experimental
observations, including the COBE data \cite{cobe}, and requirements of
self-consistency of the multi-dimensional approach to obtain limits
on parameters describing our cosmological scenario.

We consider Kaluza-Klein cosmological models with the spacetime
given by the direct product
$R \times {\cal M}^3_1 \times {\cal M}_2^d$. The manifold $R
\times {\cal M}_1^3$ represents our four-dimensional Universe
and we assume it to be of the
Friedman - Lema\^\i tre - Robertson - Walker type with
flat space hypersurfaces. The $d$-dimensional manifold
${\cal M}_2^d$ represents
the space of extra dimensions, often called internal space,
and it is assumed to
be a compact symmetric homogeneous space. We
restrict our considerations to the metrics of the form:
\[
g=-{\rm d}t \otimes {\rm d}t +a^2(t) \tilde{g} +b^2(t)\hat{g},
\]
where $\tilde{g}$ is the three-dimensional metric on ${\cal M}^3_1$ and
$\hat{g}$ is the $d$-dimensional metric on the internal space
${\cal M}^d_2$. We consider physical gravitational waves, i.e.
gravitational waves on ${\cal M}^{3}_{1}$. In our analysis we
assume that the only
spatial dependence is given by the eigentensors of the Laplacian
on ${\cal M}^3_1$ labelled by the wavenumber $n$ \cite{Ab},
i.e. we retain only the lowest (zero) mode on ${\cal M}^d_2$.
In terms of the conformal time $\eta$ the time-dependent amplitude
of the wave can be expressed as
$\nu_{n} (\eta )\equiv \mu_{n} (\eta )/f(\eta)$, where
$f(\eta) \equiv \sqrt{a^2(\eta) b^d(\eta)}$. Then $\mu(\eta )$ obeys
the following equation:
\begin{equation}
\label{2-10}
\mu_{n}''(\eta)+ (n^2-\frac{f''(\eta)}{f(\eta)}) \mu_{n}(\eta) =0.
\end{equation}
This equation was derived and studied first in the four-dimensional
cosmology, see, for example, \cite{Gr-CQG93}. In the multidimensional
case Eq. (\ref{2-10}) was considered in Refs. \cite{Ab,De-GV,GG}.

In the quantum-mechanical treatment \cite{Gr74-90} (see also
\cite{Gr-CQG93}) the
perturbation $h_{ij}$ becomes an operator. If we require the amount
of energy in each mode to be $\hbar \omega /2$, its
general expression is the following:
\begin{equation}
\label{3-10}
h_{ij}(\eta ,\tilde{x}^k) = 4\sqrt{\pi
}\frac{l_{Pl}b_{KK}^{d/2}}{f(\eta)}\frac{1}{(2\pi)^{3/2}}\int ^{+\infty
}_{-\infty}{\rm d}^3{\bf n}\sum_{s=1}^{s=2}p_{ij}^s({\bf n})
\frac{1}{\sqrt{2n}}(c_{\bf n}^s(\eta
)e^{i{\bf n \cdot \tilde{x}}}+c_{\bf n}^{s\dagger}(\eta )e^{-i{\bf n
\cdot \tilde{x}}}),
\end{equation}
where $\tilde{x}^k$ are the coordinates on ${\cal M}^3_1$.
We used the fact that the multidimensional
gravitational constant $G^{(4+d)}$ is related to the four-dimensional
one $G^{(4)}$ as $G^{(4+d)}=G^{(4)} V_{d}$ with the volume
of the internal space $V_{d}$ evaluated for $b=b_{KK}$, the present day
value of the scale factor of the internal space. The
polarization tensor $p_{ij}^s({\bf n})$ satisfies the relations:
$p_{ij}^sn^j=0$, $p_{ij}^s\delta ^{ij}=0$,
$p_{ij}^sp^{s'ij}=2\delta ^{ss'}$
and $p_{ij}^s(-{\bf n})=p_{ij}^s({\bf n})$. The time evolution of the
operator $h_{ij}(\eta ,\tilde{x}^k)$ is determined by the time
evolution of the operators $c_{\bf n}^s$ and $c_{\bf n}^{s\dagger }$ which
obey the Heisenberg equations:
\[
\frac{{\rm d}c_{\bf n}}{{\rm d}\eta } = -i[c_{\bf n},H] ,
\; \; \; \;
\frac{{\rm d}c_{\bf n}^{\dagger}}{{\rm d}\eta } = -i[c_{\bf n}^{\dagger},H] .
\]
The Hamiltonian $H$, providing a description in terms
of travelling waves, is given by:
\begin{equation}
\label{3-21}
H=nc_{\bf n}^{\dagger}c_{\bf n}+nc_{-{\bf n}}^{\dagger}c_{-{\bf n}}+
2\sigma (\eta )c_{\bf n}^{\dagger}c_{-{\bf n}}^{\dagger}+2\sigma ^*(\eta
)c_{\bf n}c_{-{\bf n}},
\end{equation}
where $\sigma(\eta )\equiv if'/(2f)$.
For $d=0$ the expressions of Ref. \cite{Gr-PRL} are recovered.
In the multidimensional case there is the second pump field $b(\eta)$
in addition to $a(\eta )$, which also appears in the four-dimensional
case, and the production of gravitons will be different.
The form of the Hamiltonian (\ref{3-21}) explicitly demonstrates
that, while the Universe expands, the initial vacuum state evolves
into a squeezed vacuum state with characteristic statistical properties
as discussed in Ref. \cite{Gr-95}. The Heisenberg equations are resolved
with the help of the standard Bogoliubov transformations:
$c_{\bf n}(\eta) =u_{n} c_{\bf n}(\eta_{0}) + v_{n} c_{\bf n}^{\dagger}
(\eta_{0})$ and similar one for $c_{\bf n}^{\dagger}(\eta)$. It follows that
then the function $\mu_{n}(\eta) \equiv u_{n} (\eta) + v_{n}^{*}(\eta)$
obeys the classical equation (\ref{2-10}).

In order to derive bounds on parameters of cosmological models
from COBE observational data we calculate the angular correlation
function for the temperature variation of the CMBR caused by the
cosmological perturbations (Sachs-Wolfe effect \cite{SW}). This
function depends only on
the angle $\delta $ between the unit vectors $e_1$ and $e_2$
pointing out in the directions of observation and can be expanded
in terms of the Legendre polynomials $P_{l}$ as follows:
\begin{equation}
\label{3-48}
<0|\frac{\delta T}{T}(e_1^k)\frac{\delta T}{T}(e^k_2)|0>
= l_{Pl}^2\sum _{l=2}^{\infty}K_lP_l(\cos \delta),
\end{equation}
where the multipole distributions $K_l$ are equal
to
\begin{equation}
\label{3-50}
K_l=(2l+1)l(l+1)[l(l+1)-2] b_{KK}^{d} \int _0^{\infty }
{\rm d}n n^2\Bigl|\int _0^{\eta _R-\eta _E}{\rm
d}w\frac{J_{l+1/2}(nw)}{(nw)^{5/2}}g_n(\eta _R-w)\Bigl|^2 .
\end{equation}
Here $\eta _E (\eta _R)$ is the time at which photons of the CMBR were
emitted (received). The factors $l_{Pl}$ and $b_{KK}$ come from the
normalization of the field operator (\ref{3-10}).
The function $g_n(\eta _R-w)$ is defined by the
formula: $g_n(\eta_R-w)\equiv [\mu _{n}/(\sqrt{2n}f)]'$.
These formulas will be used for the matter-dominated epoch, thus
$f$ will
be taken equal to $f=ab_{KK}^{d/2}$ and the factor $b_{KK}$
cancels out in the formula (\ref{3-50}). Contributions due to extra
dimensions enter through the function $\mu (\eta)$. The expressions
(\ref{3-48}), (\ref{3-50}) for the four-dimensional case were obtained
in \cite{Gr-PRL,Gr-rot}.

In this article we consider the evolution of the Universe which includes
standard radiation dominated and matter dominated stages and an epoch
during which the three-dimensional space ${\cal M}_{1}^{3}$ experiences
inflationary expansion. We choose the following scenario for the
behaviour of the scale factors:

1) Inflationary stage (I-stage):  $\eta < \eta_{1} < 0$
\begin{equation}
a(\eta)  =  l_{0}|\eta|^{1+\beta}, \; \; \;
b(\eta)  =  b_{0}|\eta|^{\gamma}. \label{Ia}
\end{equation}

2) Transition stage: $\eta_{1} < \eta < \eta_{2}$
\[
a(\eta) =  l_{0} a_{e} (\eta - \eta_{e}), \; \; \;
b(\eta) =  \left\{ \frac{B \cosh [n_{t}(\eta - \eta_{t})]}
{a_{e} l_{0} (\eta - \eta_{e})} \right\}^{2/d}.
\]

3) Radiation-dominated stage (RD-stage): $\eta_{2} < \eta < \eta_{3} $
\[
a(\eta) =  l_{0} a_{e} (\eta - \eta_{e}), \; \; \;
b(\eta) = b_{KK}.
\]

4) Matter-dominated stage (MD-stage): $\eta > \eta_{3}$
\[
a(\eta) = l_{0} a_{m} (\eta - \eta_{m})^{2}, \; \; \;
b(\eta) = b_{KK}.
\]

In order to assure the continuity of the scale factor of extra
dimensions and its first derivative between the I-stage and the
RD-stage, we have added a
period, called transition period, at which $b(\eta )$ interpolates smoothly
between its value at the end of inflation and $b_{KK}$. The
corresponding behaviour of the scale factors at the I-stage in terms of the
synchronous time is $a(t) \sim |t|^{(1+\beta)/ (2+\beta)}$, $b(t) \sim
|t|^{\gamma/(2+\beta)}$. All models with $1+\beta < 0$ ($\eta$ must
be negative in this case) describe inflationary expansion of the
three-dimensional part of the Universe. It can be shown that the case
$\beta = -2$ corresponds to the de
Sitter expansion. The cases $\beta < -2$ correspond to  power-law inflation,
i.e. $a(t) \sim t^m$, $m>1$, and the cases $-2 < \beta < -1$ correspond to
super-inflation of the type $a(t) \sim |t|^m$, $m<-1$, $t<0$.
The function $b(\eta)$ is taken to be constant for
$\eta > \eta_{2}$. This agrees with strong bounds on the time variation
of the scale factor of extra dimensions obtained in \cite{KPW-B}
at the RD- and MD-stages. Its behaviour at the transition stage mimics
a period of slowing down of the evolution of $b(\eta)$
in the process of compactification that appears in many Kaluza-Klein
cosmological models (see, for example, \cite{KRT}). At the moment we
do not impose
any restrictions on the parameter $\gamma$ characterizing $b(\eta)$
at the I-stage. Our scenario is rather general since in most of the
known models of Kaluza-Klein cosmology \cite{Kolb-book}-\cite{GSV}
the behaviour of the scale factors at the inflationary-compactification
stage is of the same type as the one described by Eqs. (\ref{Ia}).

To calculate the angular variation of the temperature of the CMBR we
need to solve Eq. (\ref{2-10}). The initial conditions on the wave
amplitude, corresponding to the vacuum spectrum of the perturbations
characterized by "a half of the quantum" in each mode, are the following:
$\mu(\eta_{0}) =1$, $\mu'(\eta_{0}) = -in$, where $\eta_{0} < 0$ is such
that $|\eta_{0}| \gg |\eta_{1}|$ \cite{Gr-CQG93}.
some $\eta _0<<\eta _1<0$. Then the exact solution of
Eq. (\ref{2-10}) is equal to

1) I-stage:
\[
\mu(\eta) =(n\eta )^{1/2} A H^{(2)}_{N+\frac{1}{2}}(n\eta ),
\]
where $H^{(2)}_{\nu}(z)$ is the Hankel function of the second kind,
$N=\beta + (\gamma d)/2$ and
$A=-i\sqrt{\pi/2}\exp [i(n\eta_0-\pi N/2)]$.

2) Transition stage:
\[
\mu(\eta) =B_{1}e^{-iw (\eta-\eta_{t}) }+
B_{2}e^{iw (\eta-\eta_{t})},
\]
where $w^{2}=n^2-n_t^2$. If $n>n_t$, $\mu(\eta)$ is an oscillatory function
whereas if $n<n_t$, $\mu $ is the sum of the exponentially growing and
exponentially decreasing solutions.

3) RD-stage:
\[
\mu(\eta) =C_{1}e^{-in(\eta -\eta _e)}+C_{2}e^{in(\eta -\eta _e)}.
\]

4) MD-stage:
\begin{equation}
\mu(\eta) =\sqrt{\frac{\pi
z}{2}}\biggl(D_{1} J_{\frac{3}{2}}(z)+D_{2} J_{-\frac{3}{2}}(z)\biggr),
\label{MDmu}
\end{equation}
where $z\equiv n(\eta -\eta _m)$. The coefficients $B_{i}$, $C_{i}$ and
$D_{i}$ $(i=1,2)$ are determined by matching the solution and its first
derivative.

To set the
scale for $\eta$ it is convenient to choose $\eta_{R}- \eta_{m} = 1$.
All realistic cosmological models should give
$a(\eta_{E})/a(\eta_{R}) \approx 10^{-3}$,
$a(\eta_{3})/a(\eta_{R}) \approx
10^{-4}$ and $a(\eta_{1})/a(\eta_{R}) = k$, where
$3 \cdot 10^{-32} < k < 3 \cdot 10^{-12}$. The lower
bound on $k$ corresponds to the case when the radiation dominated
expansion of the three-dimensional part of the Universe starts at the
Planckian energy densities, whereas the upper one corresponds to
the case when this process starts at the nuclear energy densities.
{}From the continuity conditions on the scale factors $a(\eta)$
and $b(\eta)$ and its first derivatives $a'(\eta)$ and $b'(\eta)$
we obtain the following expressions for the parameters of the scenario
in terms of $\beta$ and $k$:
\begin{eqnarray}
\eta_{1} &=& 50k(1+\beta ), \nonumber  \\
\eta_{3} &=& 50k\beta +0.5\cdot 10^{-2}, \; \; \;
\eta_{e} = 50k\beta, \; \; \;
\eta_{m} = -0.5\cdot 10^{-2}+50k\beta, \nonumber  \\
a_{e} &=& -(1+\beta )(50k|1+\beta |)^{\beta }, \; \; \;
a_{m} = 50|1+\beta |(50k|1+\beta |)^{\beta }.  \nonumber
\end{eqnarray}
The characteristic scale $l_{0}$ in Eq.
(\ref{Ia}) is given by the relation
\begin{equation}
\frac{l_{Pl}}{l_{0}} = 25 \left( \frac{l_{Pl}}{l_{H}} \right)
     (50k)^{\beta} |1+\beta|^{(1+\beta)},    \label{lPll0}
\end{equation}
where $l_{H} \equiv a^{2}(\eta_{R})/a'(\eta_{R})$ is the present day
Hubble radius. We take it to be equal to $l_{H}=10^{61} l_{Pl}$.

It turns out that in order to get relations on the parameters
$\eta_{2}$, $\eta_{t}$, $B$ and $n_{t}$ one has to solve the
transcendental equation
\begin{equation}
\tanh \left( \sqrt{r^2(x^2-p^2)+1}-x+\arg\tanh \frac{p}{x}\right)=
\frac{1}{\sqrt{r^2(x^2-p^2)+1}},    \label{transeqn}
\end{equation}
where $p=(N+1)/(\beta+1)$ and $r=(b(\eta_{1})/b_{KK} )^{d/2}$.
This equation arises from the continuity conditions on $b(\eta)$ and
$b'(\eta)$. If $x(r,p)$ is a solution of Eq. (\ref{transeqn}) for given
$r$ and $p$ we get
\begin{eqnarray}
n_{t} & = & \frac{1+\beta}{\eta_{1}} x(r,p) = \frac{x(r,p)}{50k},
\; \; \; B =  l_{0} b_{0}^{d/2} \frac{\sqrt{x^{2}(r,p)-
p^{2}}}{x(r,p)} |\eta_{1}|^{N+1},     \label{nt}  \\
\eta_{2} - \eta_{1} & = &
   50k \frac{\sqrt{r^{2}(x^{2}(r,p)-p^{2})+1} - x(r,p)}{x(r,p)},
\; \; \; \eta_{1} - \eta_{t}  =  \frac{1}{n_{t}} \arg
   \tanh \frac{p}{x(r,p)}.     \label{eta2-eta1}
\end{eqnarray}
{}From Eqs. (\ref{nt}) and (\ref{eta2-eta1}) we see that physically
acceptable solutions must satisfy $x(r,p) > |p|$.
It can be shown that such solutions exist for $r>1$ and
$p < p_{crit}(r) < 1/r$ only. For example, for $r=2$ $p_{crit}=0.22$
and if we choose $p=0.1$, the solution is $x=0.373$. From eq.
(\ref{eta2-eta1}) it follows that $n_{t}(\eta_{2}-\eta_{1}) = 0.907$
then.

Thus, the consistency of the scenario requires $p<1$. This means
that $\gamma>0$ and the scale factor of the internal space decreases
at the I-stage. The condition $r>1$ means that this scale factor
continues to decrease at the transition stage. Since we
assume that the classical evolution of the multidimensional Universe may start
at some value of $b(\eta_{0})$ close to the Planck length, it is reasonable to
limit our consideration to not very big values of $r$, otherwise $b_{KK}$
appears to be much smaller than $l_{Pl}$. We restrict ourselves to the range
$1 < r < 10$.

To impose further restrictions let us consider a condition on the size
of the space of extra
dimensions after the end of the compactification process.
It is easy to show that
\begin{equation}
\frac{l_{Pl}}{b_{KK}}  =  \left( \frac{l_{0}}{b_{0}} \right)
\frac{r^{2/d}}{4} S, \; \; \; \mbox{with } \; \; \;
S  =  100 \left( \frac{l_{Pl}}{l_{H}} \right)
   |1+\beta|^{1+\beta - \gamma} (50k)^{\beta - \gamma}. \label{S-defn}
\end{equation}
We make a natural supposition that at the beginning of the
inflation all space dimensions in the early Universe were of
the same order. In this article for the sake of simplicity we take
$l_{0}=b_{0}$.

The only experimental bound on the size of $b_{KK}$ comes from the fact
that no effects of extra dimensions are observed in high energy
particle experiments. This, apparently, tells us that $\hbar c/b_{KK} >
(1 \div 10) $ TeV. On the other hand the classical description of the
background dynamics can be trusted only if $b_{KK}$ is not much
smaller than $l_{Pl}$. These arguments imply that $S$ in eq.
(\ref{S-defn}) should belong to the interval $10^{-16}<S<1$ that
gives certain restrictions on $\beta$ and $\gamma$.
However, we found that it is more convenient to analyze these
restrictions together with the bounds coming from the COBE experiment.
We are going to derive these bounds right now.

It can be shown that the
main contribution to the multipole distributions $K_{l}$ of the
correlation function for the temperature variation of the CMBR , Eqs.
(\ref{3-50}), is given by long waves, namely by the waves with the
wavelength equal or larger than the present day Hubble radius. Such
waves have wavenumbers $n \ll n_{H}= 4 \pi$. The approximate form of
the solution for these wavenumbers is the following:
\begin{equation}
\mu_{n}(\eta) \approx - i e^{in\eta_{0}} 50 \frac{\Psi (N)}{r}
  (1+\beta)^{1+N} (50k)^{N} n^{1+N} (\eta - \eta_{m})^{2},
    \label{mu-appr}
\end{equation}
where $\Psi (N) \equiv \sqrt{\pi /2} \exp (i\pi N/2) [\cos (N\pi )
2^{N+1/2}\Gamma (N+3/2)]^{-1}$.

It can be shown that for the separation angle $\delta = 0$ the variance
of $\delta T/T$ can be approximately characterized by
\[
<0|\frac{\delta T}{T}(e^k)\frac{\delta T}{T}(e^k)|0> \sim
10^{-5}  h^{2}_{H},  \]
where $h_{H}$ is the characteristic spectral component defined by
$ h(n;\eta) = l_{Pl} n |\mu (\eta)|/a(\eta)$ and evaluated at
$n=n_{H}$, $\eta = \eta_{R}$ (see the discussion in Ref. \cite{Gr-95}).
Using Eqs. (\ref{lPll0}) and (\ref{mu-appr}) we obtain that
\begin{equation}
 h(n;\eta_{R}) = 25 \left( \frac{l_{Pl}}{l_{H}} \right)
  \frac{|\Psi (N)|}{r} |1+\beta|^{(1+N)} (50k)^{N} n^{2+N},
  \label{h-final}
\end{equation}
In this expression the limit
of the four-dimensional case is achieved by putting $d= 0$. Then
$N=\beta$, $r=1$ and the analogous formula of ref. \cite{Gr-PR93}
is recovered. COBE experimental results give $(\delta T/T)_{exp} \approx
6 \cdot 10^{-6}$ \cite{cobe}, consequently $h_{H}$ must be
of the order $10^{-4}$.

Thus, we have two conditions to be satisfied:
\begin{equation}
  h_{H} = 10^{-4} \; \; \; \mbox{and} \; \; \;
  10^{-16} \leq  S \leq 1,              \label{hScond}
\end{equation}
where $S$ and $h(n_{H})$ are given by Eqs. (\ref{S-defn}) and (\ref{h-final})
respectively. Resolving these conditions we obtain
\begin{eqnarray}
N &=& \frac{53}{3+\lg (k/3)},  \label{N-bound} \\
\frac{d+2}{2} \gamma &=& - \frac{71 + 6 \lg (k/3)}{(3+\lg (k/3))
    (2+\lg (k/3))} - \frac{\lg S}{2+\lg (k/3)}.  \label{gamma-bound}
\end{eqnarray}
When $k/3$ varies within the interval $10^{-32} < k/3 < 10^{-12}$,
$N$ changes within the bounds $-5.9 < N < -1.8$, which essentially
coincide with the bounds on $\beta$ coming from the analogous
condition in the four-dimensional scenario \cite{Gr-PR93}. In
addition, one should check that the mean square value of the
field is finite in the limit of small wave numbers $n$. This gives
additional restriction $N > -2$, see \cite{Gr-PR93}.
{}From Eq. (\ref{gamma-bound}) we conclude that upper bound on this
parameter is given by $(d+2)\gamma/2 < 0.14$.
Recall that for the scenario to be consistent $\gamma$ must
be positive.

The region of allowed values of $\beta$ and $\gamma$ for $d=6$
is presented in Fig. 1. For other $d$ the shape of the region remains
the same, however its area decreases when $d$ grows.
The consistent values of $S$ within our
scenario, i.e. those which admit positive values of $\gamma$, belong
to the interval $10^{-4} \leq S \leq 1$. We would like to emphasize that
$S=10^{-4}$ corresponds to $\hbar c/b_{KK} = 10^{15}$ GeV, which is
approximately the scale of the Grand Unification.

We are unaware of any model of Kaluza-Klein cosmology which agrees with
the limits on $\beta$ and $\gamma$ obtained above. For example,
among the models corresponding to our scenario, one finds that
$\beta=-5/4$, $\gamma = 1/4$ for $d=6$ in the perfect-fluid-dominated
model \cite{ABE}, $\beta=-1.26$, $\gamma=0.22$ in the $D=4+d=11$
supergravity
with toroidal compactification \cite{MN}, $\beta=-14/11$, $\gamma=1/11$
for $d=22$ in the model of string-driven inflation \cite{GSV}. It is
easy to check that none of these models satisfy the bounds.
Results of more complete analysis of Kaluza-Klein models will be
presented elsewhere.

We would like to mention that using the exact solution for
$\mu (\eta)$ and the formula (\ref{3-50}) we can compute
the multipole distributions $K_l$
contributing to the angular correlation function, Eq.
(\ref{3-48}). However, it can be shown that the ratios $K_{l}/K_{2}$
are the same as in the four-dimensional case considered in \cite{Gr-PR93}
provided we replace $\beta $ with $N$.  Therefore, it seems that
multidimensional cosmological models satisfying the first condition in
Eq. (\ref{hScond}) also give the values of $K_{l}$ which roughly
agree with the experimental data.

The transcendental equation
(\ref{transeqn}) requires $\gamma $ to be positive, that means
that the size $b$ of the internal space decreases at
the I-stage. Therefore, though the effect of the transition
period on the amplitude $\mu (\eta )$ is small, we see that the
range of the variation for the
parameter $\gamma $, allowed by the background model,
depends on the details of the transcendental equation.
It is quite possible that
for other types of the behaviour of the scale factors during
the transition period the case $\gamma < 0$ is permitted. Our
analysis must be re-examined for that case. We will consider
this possibility in a future publication.

In this article we considered the contribution to the angular variation
in the temperature of the CMBR coming from the gravitational wave
perturbations generated quantum-mechanically as a result of parametric
interaction of the perturbations with strong variable
background gravitational fields in the Early Universe. Due to
universal character of this mechanism such wave perturbations have
been generated inevitably, hence firmness of the
limits (\ref{N-bound}) and (\ref{gamma-bound}) on $\beta$ and $\gamma$,
which are the main results of our paper. Contributions due to
perturbations of other types (density and rotational perturbations,
non-zero modes on ${\cal M}_2^d$) in the Kaluza-Klein cosmology should be
analyzed as well, however they are beyond the scope of the present article.

The limits (\ref{N-bound}) and (\ref{gamma-bound})
(and their graphical representation in Fig. 1)
show that the conditions of consistency
and the recent data from the COBE experiment leave a room for
multidimensional cosmological models as candidates for the description
of the Early Universe. However, the limits are rather restrictive and
we did not find any concrete model satisfying them. Taking into account
further restrictions imposed, for example, by the pulsar-timing data or
by the future LIGO experiment will allow to make the limits on
multidimensional models more restrictive thus questioning the very validity
of the Kaluza-Klein hypothesis. We would like to mention that the
spectral energy density of the gravitational waves and
some observational bounds on multidimensional cosmological
models were studied in Ref. \cite{GG}.

\vspace{1cm}

\noindent{\large \bf Acknowledgments}

It is a pleasure for us to thank Leonid Grishchuk and Pedro Pascual
for valuable discussions and Richard Kerner for constant encouragement
and usefull comments. J.M. would
like to thank the Universitat de Barcelona for warm hospitality, the
Minist\`ere de la Recherche et de l'Enseignement Sup\'erieur for a research
grant. Yu.K. would like to thank the Laboratoire de Gravitation
et Cosmologies Relativistes, Universit\'e Pierre et Marie Curie for warm
hospitality. This investigation has been supported by
M.E.C. (Spain), grant SAB94-0087, and by CIRIT (Generalitat de
Catalunya).

\newpage

\section*{Figure caption}

\begin{description}
  \item[Fig. 1] The region of values of the parameters
                $\beta$ and $\gamma$ given by the equations (\ref{N-bound})
                and (\ref{gamma-bound}) for $d=6$. The four presented
                curves correspond to various values of $y= \lg S$. The
                dashed straight line is given by $N=-2$. The hatched
                region is the region of values of $\beta$ and $\gamma$
                allowed by the observational data and the consistency
                conditions discussed in the article.

\end{description}

\end{document}